\documentclass[aps,prl,twocolumn,superscriptaddress]{revtex4}
\usepackage{amsmath,amssymb,color}
\usepackage{graphicx}
\usepackage{bm}

\def\be{\begin{equation}}
\def\ee{\end{equation}}
\def\bmo{Ba$_3$Mn$_2$O$_8$}

\def\BaFS{$^{135,137}$Ba}

\def\HparC{$\mathbf{H}\parallel\mathbf{c}$}
\def\HperpC{$\mathbf{H}\perp\mathbf{c}$}
\def\HcOne{$H_{c1}$}

\def\c{$\mathbf{c}$}
\def\Mlpar{$M_{\ell\parallel}$}
\def\Mlperp{$M_{\ell\perp}$}

\begin{document}

\title{Critical properties of the $S$=1 spin dimer compound
Ba$_3$Mn$_2$O$_8$}

\author{S.~Suh}
\affiliation{Department of Physics and Astronomy, UCLA, Los Angeles, CA 90095-1547 USA}
\author{K.~A.~Al-Hassanieh}
\affiliation{Theory Division, Los Alamos National Laboratory, Los Alamos, NM 87545 USA}
\author{E.~C.~Samulon}
\affiliation{Geballe Laboratory for Advanced Materials and Department of Applied
Physics, Stanford University, Stanford, CA 94305 USA}
\author{J.~S.~Brooks}
\affiliation{Department of Physics and National High Magnetic Field Laboratory,
Florida State University, Tallahassee, FL 32310 USA}
\author{W.~G.~Clark}
\affiliation{Department of Physics and Astronomy, UCLA, Los Angeles, CA 90095-1547 USA}
\author{P.~L.~Kuhns}
\affiliation{Department of Physics and National High Magnetic Field Laboratory,
Florida State University, Tallahassee, FL 32310 USA}
\author{L.~L.~Lumata}
\affiliation{Department of Physics and National High Magnetic Field Laboratory,
Florida State University, Tallahassee, FL 32310 USA}
\author{A.~Reyes}
\affiliation{Department of Physics and National High Magnetic Field Laboratory,
Florida State University, Tallahassee, FL 32310 USA}
\author{I.~R.~Fisher}
\affiliation{Geballe Laboratory for Advanced Materials and Department of Applied
Physics, Stanford University, Stanford, CA 94305 USA}
\author{S.~E.~Brown}
\affiliation{Department of Physics and Astronomy, UCLA, Los Angeles, CA 90095-1547 USA}
\author{C.~D.~Batista}
\affiliation{Theory Division, Los Alamos National Laboratory, Los Alamos, NM 87545 USA}

\date{\today}

\begin{abstract}
\bmo\ is a hexagonally coordinated  Mn$^{5+}$ $S$=1 spin dimer system with small
uniaxial single-ion anisotropy. \BaFS\ NMR spectroscopy is used to establish the lower
critical field $H_{c1}$ of distinct field-induced phases for \HparC,\HperpC, and measure the longitudinal ($M_\ell$) and transverse ($M_t$) magnetizations in the vicinity of the quantum critical point (QCP).
$M_{\ell\parallel}(T,H_{c1})$, $M_{\ell\perp}(T,H_{c1})$ are reproduced by solving a
low-energy model for a dilute gas of interacting bosons. $M_{t\parallel}(T\to0,H=H_{c1})$
($M_{t\perp}(T\to0,H=H_{c1})$) follows the expectation for a  BEC (Ising-like) QCP.\\

PACS nos. 75.45.+j,75.40.Cx,75.40.-s,76.60.Cq

\end{abstract}


\maketitle

Recent investigations of field-induced phases in $S$=1/2 magnetic insulators typify the opportunities for studying the problem of Bose Einstein condensates (BEC's) specifically \cite{Giamarchi:2008}, and quantum criticality more generally. In spin-dimer, and other spin-gapped systems, the ground state is a singlet while the lowest excited states are a mode of triplet excitations \cite{Nikuni:2000,Jaime:2004}. The magnetic field tunes the chemical potential for triplet excitations through zero at the critical field $H_{c1}$ producing a controlled density of triplets, that can either condense or crystallize into a superlattice depending on the balance between kinetic and potential energies \cite{Kodama:2002,Rice:2002}. The Hamiltonian has U(1) rotational symmetry in the idealized case, and this symmetry is spontaneously broken in the condensed phase with the development of a finite transverse magnetization $M_t$.

\begin{figure}[htb]
\includegraphics[width=3in]{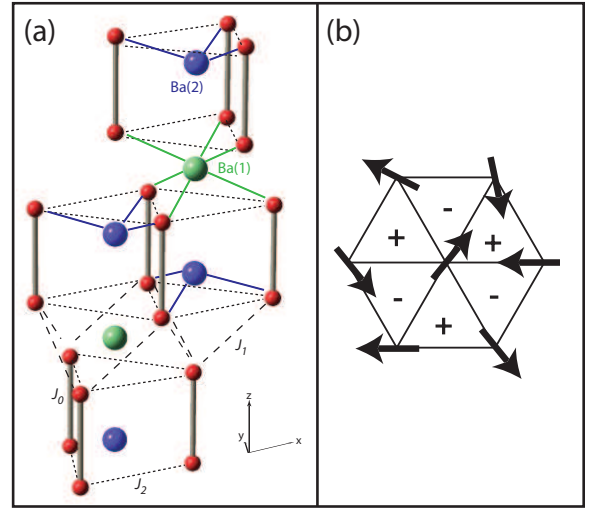}
\caption{(color online) a) Ionic arrangement of \bmo\ showing location of Ba sites relative to the Mn ions; oxygen ions are not shown. The exchange couplings are indicated by solid ($J_0$) and dashed ($J_1$, $J_2$) lines. b) Illustration of the two chiralities, with the transverse spin directions denoted by the arrows, with $\alpha=129^\circ$ (see text).}
\label{fig:BMOstructure}
\end{figure}

From what is known about the spin-dimer system \bmo\ \cite{Weller:1999}, these conditions hold for \HparC\ \cite{Samulon:2008}. However, the evolution of the phases in a magnetic field
is known to deviate from the simplest  $S$=1/2 isotropic case in a number of ways \cite{Uchida:2002,Tsujii:2005,Samulon:2008}. Two magnetization plateaus with $\langle S_z \rangle$=1 (per dimer) and $\langle S_z \rangle$=2 are observed as a result of the $S$=1 state of the Mn$^{5+}$ ions \cite{Uchida:2001,Uchida:2002}. In addition, a small single-ion uniaxial anisotropy is understood to produce new boundaries in the ordered phases for $\mathbf{H}$ tilted from the \c-axis. While this anisotropy is not relevant for \HparC, its influence is most prominent for \HperpC, where there is evidence for an additional phase II, stabilized only near $H_{c1\perp}$ and the other three critical fields. Further, the hexagonal coordination of the layers leads to geometric frustration. The near-neighbor transverse spin components would be rotated by $\alpha$=120$^\circ$ in an isolated triangular layer \cite{Singh:1992,Kawamura:1998}. However, interlayer coupling (Fig. \ref{fig:BMOstructure}) leads to $\alpha\to120^{\circ}+\epsilon$ with $\epsilon\sim 9^\circ$), because incommensurate spin ordering partially releases the interlayer frustration.

Presented here are results of $^{135,137}$Ba NMR spectroscopy studies in the high symmetry phase near \HcOne, and in the ordered phases I (\HparC, \HperpC) and II (\HperpC).  The NMR shifts are used to establish the magnetization as a function of temperature at $H=$\HcOne. For the case \HparC, the result is consistent with the expectations for a BEC-QCP, {\it i.e.}, $M(T\to 0,H_{c1})\sim T^{1.5}$. Both the universal and the non-universal ($T>$100mK) regimes are quantitatively described using an effective low-energy theory for a dilute gas of bosons. Quantitative differences are observed for \HperpC, in agreement with the expectation for the Ising-like (Z$_2$) broken symmetry of phase II \cite{Samulon:2008}. Further, we investigate the longitudinal and transverse magnetization ($M_l$, $M_t$) of the ordered phases, and establish that the field dependence of the transverse magnetization follows the expected mean-field behavior ($D=d+z \geq 4$). We also conclude that the line of transitions dividing I/II is discontinuous.

The maximum temperature of the ordered phases is $T_m\equiv$0.9K \cite{Uchida:2002}, so the measurements reported here were performed on a single crystal placed inside the mixing chamber of a dilution refrigerator. \BaFS\ ($^{135,137}$I=3/2) NMR spectroscopy was performed in magnetic fields $H\le$120kOe using a top-tuned configuration. The platform holding the sample and coil is rotated by an Attocube piezoelectric motor. At the higher fields available at the NHMFL, we used a bottom-tuned $^3$He system. The diagonal hyperfine couplings were determined by comparing high temperature measurements of the shift ($T\ge$20K) to susceptibility measurements \cite{Uchida:2002}. Orbital and quadrupolar couplings were determined from the shifts measured at the lowest temperatures for $H<H_{c1}$. Some of the NMR parameters are summarized in Table \ref{table:NMRparameters}.
\begin{table}[b]
\begin{tabular*}{3in}{@{\extracolsep{\fill}}|c||c|c|c|}
\hline
 & $A_{aa}$ & $A_{cc}$ & $^{137}\nu_Q$ (MHz) \\
\hline\hline
  Ba(1) & 0.26(1) & 0.35(2) & 54.7(2) \\
\hline
  Ba(2) & 0.18(1) & 0.11(1) & 10.8(2) \\
\hline
\end{tabular*}
\caption{Selected NMR parameters for the two sites shown in Fig. \ref{fig:BMOstructure}.
$\nu_Q\equiv e^2qQ$, with $^{135}Q$ ($^{137}Q$)=0.18 (0.28) $\times 10^{-24}$
cm$^2$. The hyperfine coupling constants are reported in $\mu_B$/Mn$^{5+}$.}\label{table:NMRparameters}
\end{table}

The $S_z=1$ triplet excitations become gapless at $H_{c1}$, while the other two
triplets and the quintets have a gap of order $J_0$. Therefore, only $S^z=1$ triplets can be created at low energies for $H \sim H_{c1}$. These triplets are hard-core bosons with an effective chemical potential $\mu =g_{\nu \nu} \mu_B (H-H_{c1})$ ($\nu=\{ a, b, c \}$). The effective Hamiltonian for ${\bf H} \parallel
{\bf c}$ ($\nu=c$) or ${\bf H} \perp {\bf c}$ ($\nu=a,b$) is:
\begin{eqnarray}
{\cal H} &=& \sum_{\bf q} (\epsilon_{\bf q} - \mu) b^{\dagger}_{\bf q} b^{\;}_{\bf q}
+(1 - \delta_{\nu,c}) \frac{g {\cal J}_{\bf q}}{2} \sum_{\bf q}  (b^{\dagger}_{\bf q}
b^{\dagger}_{-\bf q}
+ b^{\;}_{\bf q} b^{\;}_{-\bf q})
\nonumber \\
&+& \frac{v_0}{2N} \sum_{\bf q, k, k'}
b^{\dagger}_{\bf k} b^{\dagger}_{\bf k'} b^{\;}_{\bf k' - q} b^{\;}_{\bf k + q},
\end{eqnarray}
where $\epsilon_{\bf q} = \sqrt{\Delta_1^2 + \frac{8}{3} \Delta_1 {\cal J}_{\bf q}}-
\sqrt{\Delta_1^2 + \frac{8}{3} \Delta_1 {\cal J}_{\bf Q}}$,
${\bf Q}$ is the wave-vector that minimizes $\epsilon_{\bf q}$,
$\alpha_{\bf q}= g {\cal J}({\bf Q})$,
$\Delta_1 =1.65$meV is the single-dimer singlet to $S^z=1$ triplet gap,
${\cal J}_{\bf q} = 2(J_2 - J_3) \gamma^2_{\bf q} + \frac{J1}{2} \gamma^1_{\bf q} +
\frac{J4}{2} \gamma^3_{\bf q}$, and
\begin{eqnarray}
\gamma^1_{\bf q} &=& \cos{q_3}+\cos{(q_3-q_1)}
+\cos{(q_3-q_2)},
\nonumber \\
\gamma^2_{\bf q} &=& \cos{q_1}+\cos{q_2}
+\cos{(q_1-q_2)},
\nonumber \\
\gamma^3_{\bf q} &=& \cos{(q_3 -q_2 +q_1)}
+\cos{(q_3 -q_1 +q_2)}
\nonumber \\
&+&\cos{(q_3 -q_1 -q_2)}.
\end{eqnarray}
The relative magnitude of the exchange anisotropy $g=0.088$ was obtained by fitting the
difference between the values of $H_{c1 \parallel}$ and
$H_{c1 \perp}$: $\Delta H_{c1}=89.3-86.4=2.9$kOe.

The exchange constants \cite{Stone:2008a} and the g-factors are $J_1=0.118$meV, $J_2-J_3=0.114$meV, $J_4=0.037$meV,
$g_{cc}=1.98$ and $g_{aa}=1.97$. The effective repulsive interaction
$v_0= \Gamma_{\bf 0}({\bf Q},{\bf Q})$ results for summing the ladder diagrams for the bare interaction vertex  $V_{\bf q}$ \cite{Beliaev:1958}:
\begin{equation}
\Gamma_{\bf q}({\bf k},{\bf k'}) = V_{\bf q} - \int_{-\pi}^{\pi} \frac{d^3p}{8\pi^3}
V_{\bf q - p} \frac{\Gamma_{\bf  p}({\bf k},{\bf k'})}{\epsilon_{\bf k+p} +
\epsilon_{\bf k'-p}}
\label{ladder}
\end{equation}
For Ba$_3$Mn$_2$O$_8$, we have $V_{\mathbf q} = U + (J_2 + J_3) \gamma^2_{\bf q} +
\frac{J1}{2} \gamma^1_{\bf q} +
\frac{J4}{2} \gamma^3_{\bf q}$, where $U \to \infty$ comes from the hard-core repulsion, while the rest of
the terms correspond to the off-site repulsive interactions the result from the Ising terms of the inter-dimer
exchange couplings. By solving Eq.(\ref{ladder}), we obtain $v_0=0.9$meV for $J_2+J_3=2.82$K.
The value of $J_2+J_3=$ is obtained by fitting $H_{c2}\simeq 27$T for
${\bf H} \parallel {\bf c}$. We note that the second term of ${\cal H}$ breaks the U(1) symmetry
associated to the conservation of the total number of bosons ($M_l$) for ${\bf H} \perp {\bf c}$.
This term is originated by the effective exchange anisotropy found in Ref. \cite{Samulon:2008}.
Consequently, we expect an Ising-like (broken Z$_2$) quantum phase transition (QPT)
for ${\bf H} \perp {\bf c}$ in contrast to the BEC-QPT that occurs for ${\bf H} \parallel {\bf c}$.

\begin{figure}[htb]
\includegraphics[width=3in]{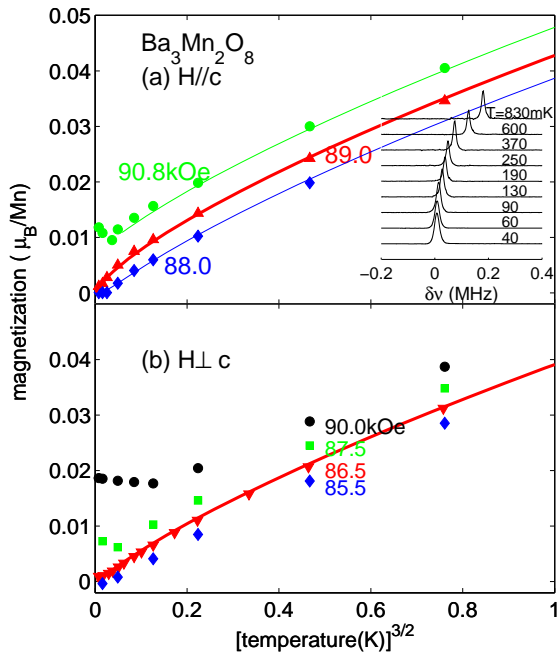}
\caption{(color online) ({\it a}) $M_{l\parallel}$ {\it vs.} $T$ for selected
magnetic fields close to $H_{c1\parallel}$. The inset shows a sequence of spectra
recorded at different temperatures. The solid and dashed curves are from ladder
diagram calculations (see text). ({\it b}) The same as ({\it a}), for
$\mathbf{H}\perp\mathbf{c}$.}
\label{fig:MagAtHc1}
\end{figure}

Magnetization results for \HparC\ are shown in Fig. \ref{fig:MagAtHc1}a for several fields near $H_{c1}$.
The curve measured at $H=H_{c1}=89.3$kOe is consistent with the expectation \Mlpar$\sim T^{3/2}$ for  $T \to 0$. The red line is the result  of a  Hartree-Fock decoupling of the last term of ${\cal H}$ that in the
disordered phase has the effect of renormalizing the chemical potential $\mu_{\rm eff}= \mu -2 v_0 \rho$
($\rho$ is the density of bosons) \cite{Nikuni:2000}. Calculations for field values differing from
$H_{c1}$, $H=H_{c1}-1.3$kOe (blue) and $H=H_{c1}+1.5$kOe (green), also match the NMR shift data well.
There is a $20\%$ disagreement if only the hard-core repulsion is included in Eq. (\ref{ladder}).

When the applied field is rotated to the $ab$ plane, the ordered phase II bordering
the paramagnet is believed to be Ising-like, with transverse spins
confined to the $\mathbf{c}$ direction. In this case, we expect \Mlperp$\sim T^2$ for $T \to 0$ and $H=H_{c1\perp}$. The NMR results in Fig. \ref{fig:MagAtHc1}b agree well with the mean field treatment of
${\cal H}$. The anisotropy term also has the effect of lowering the critical field $H_{c1}$.
In confining the transverse spins to the $\mathbf{c}$-axis, the energy gain associated with
the broken symmetry is reduced slightly, and consequently \Mlperp$<$\Mlpar. The outcome is
consistent with the anisotropy parameter $D$=32$\mu$eV as established by electron paramagnetic resonance \cite{Hill:2009}.

\begin{figure}
\includegraphics[width=3in]{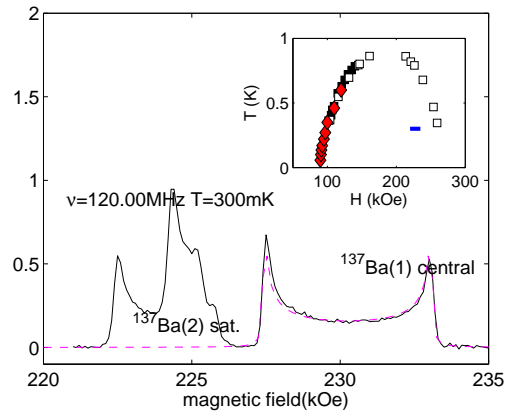}
\caption{(color online) Field-swept spectra of $^{137}$Ba(2) satellite and
$^{137}$Ba(1) central transition. The dashed line is the expectation for a plane
wave incommensurate hyperfine field modulation. The partial T/H phase
diagram in the inset results from NMR relaxation data
(solid red diamonds) \cite{Suh:2009a} and thermodynamic measurements (open squares)
in Ref. \cite{Samulon:2008}. The blue line marks the temperature and field range for the main panel.}\label{fig:Ba1Ba2spectra}
\end{figure}

\begin{figure}
\includegraphics[width=3in]{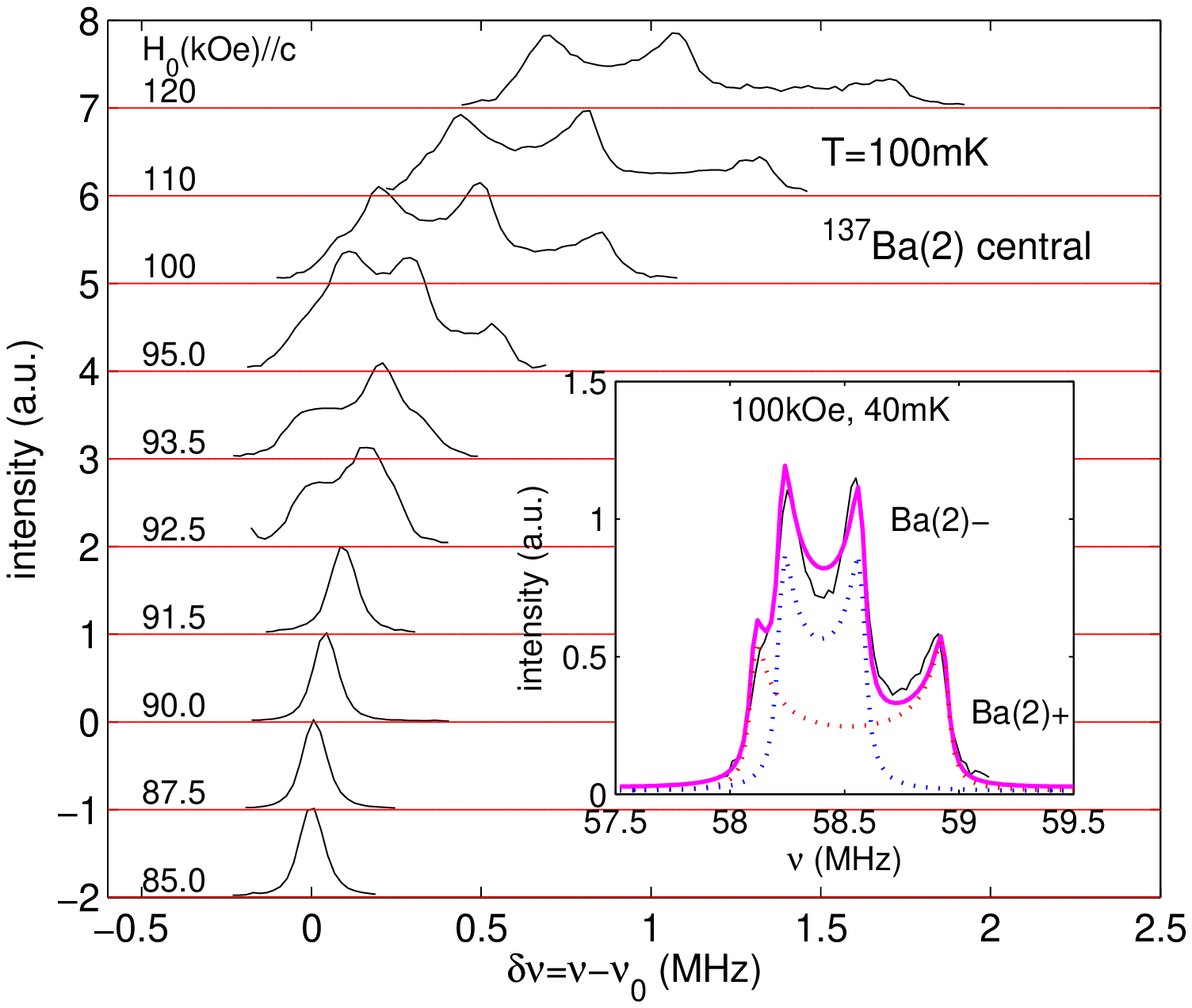}
\includegraphics[width=3in]{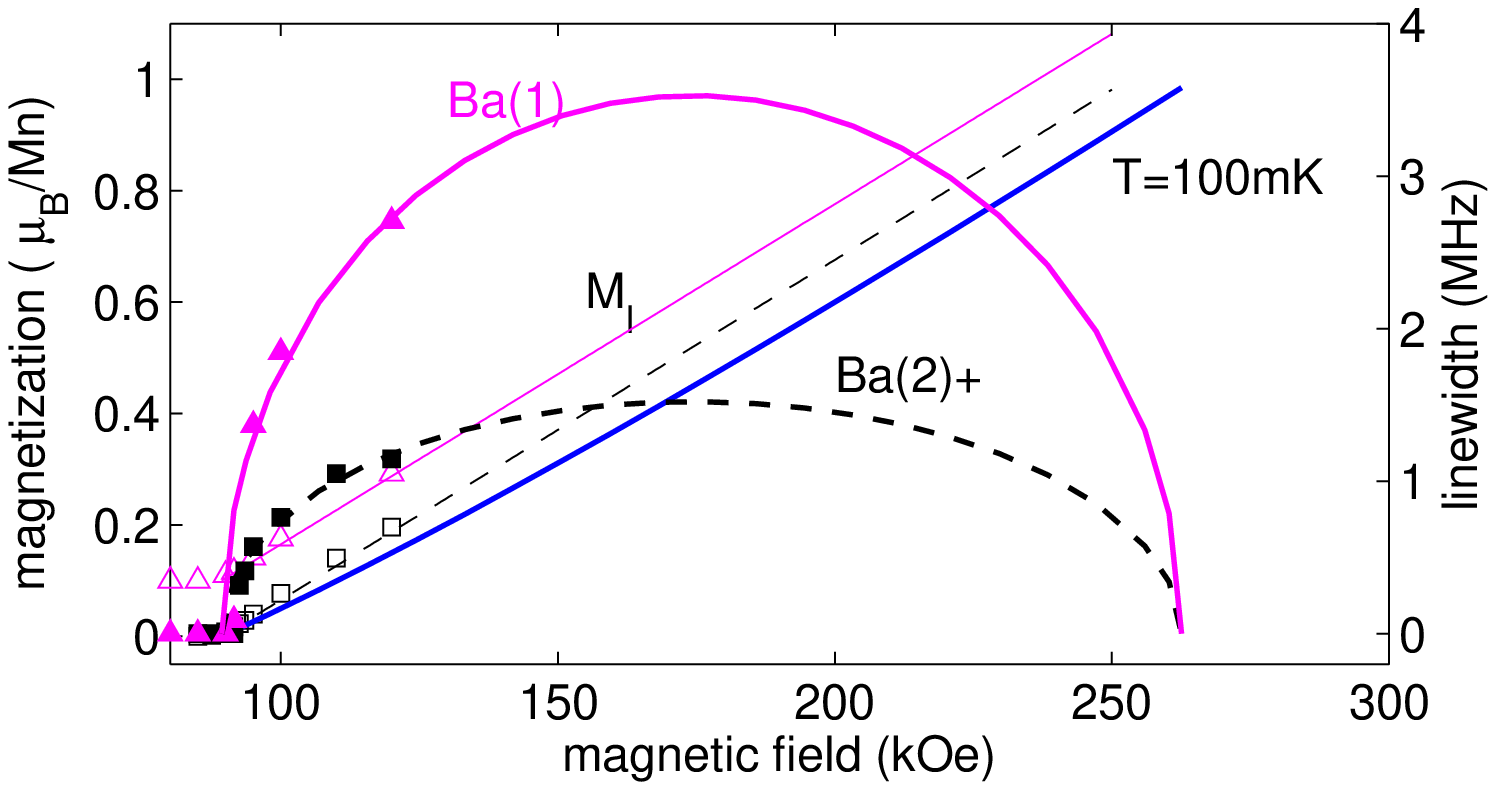}
\caption{(color online) a) Frequency-swept spectra for Ba(2) site near to $H_{c1}$.
The inset shows a comparison of the result to two independent plane wave
incommensurate modulations of the hyperfine field. The two modulations correspond to
Ba(2) locations of positive (+) and negative (-) chiralities. b) Longitudinal and
transverse magnetization inferred from the spectra.}\label{fig:Ba2HparCnearHc1}
\end{figure}

Turning to the transverse magnetization, Fig. \ref{fig:Ba1Ba2spectra}
shows two field-swept spectra in the condensed phase for \HparC.
The blue segment in the inset shows the location in the phase diagram
where these spectra were recorded. The spectrum of the Ba(1) site
is well reproduced by assuming a simple plane wave incommensurate modulation of the
longitudinal field  (dashed red line). The functional form is independent of
${\bf H}$ for \HparC; only the spectral width and overall shift vary.
The Ba(2) spectra are composed of two parts of equal intensity, and there appears
to be an asymmetry in the line shape. The asymmetry is reduced for fields close to $H_{c1}$
(inset of Fig. \ref{fig:Ba2HparCnearHc1}), so the lineshapes are modelled
by plane wave incommensurate structures with different widths and shifts.
A small systematic error is introduced by varying magnetic field
rather than frequency, so all subsequently shown spectra were obtained by sweeping the frequency at constant $H$.

The incommensurate modulation arises from the interlayer coupling and
the off-diagonal hyperfine coupling makes its detection possible.
In comparing the broadening of central and satellite transitions, the magnetic origin of the broadening
is confirmed \cite{Suh:2009a}. The two inequivalent sites for Ba(2) are associated
with triangular coordination of the Mn sites with opposing chirality ($\pm$)
of the transverse order. In the Ba(1) case, the point group symmetry is ($C_{3i}$): each Ba(1) site is situated equidistant from three Mn sites in each of two dimer layers. The two Mn triangles have opposite chirality in the condensed phase, and by symmetry there is no distinction of the Ba(1) sites except for the phase modulation resulting from the incommensurability. The Ba(2) sites have lower (C$_{3}$) symmetry, with the three nearest-neighbor Mn forming a triangle of specific chirality.

The spectrum broadens and shifts near $H_{c1}$.
Fig.\ref{fig:Ba2HparCnearHc1}a shows the frequency-swept Ba(2) spectra for the central (1/2$\leftrightarrow$-1/2) transition, collected for a sequence of fields close to $H_{c1}$. The field dependence of $M_t(H)$ is generated by using the model of an
incommensurate modulation of the hyperfine field (see Fig. \ref{fig:Ba2HparCnearHc1}b).
The results obtained from both Ba(1) and Ba(2) sites are included in the figure.
The solid and dashed lines are guides based on the approximate empirical relationships,
$M_l  \approx  \frac{g\mu_B(H-H_{c1})}{H_{c2}-H_{c1}}$
$M_t  =  [M_l(1-M_l)]^{1/2}.\label{eq:Mt}$
The first expression linearly interpolates between the critical fields, while the second
is the simplest (two level approach) mean field result expected for the condensed phase I.
We note that the mean field exponents, $M_l  \propto (H-H_{c1})$ and $M_t  \propto (H-H_{c1})^{1/2}$,  are
correct because the effective dimension $D=d+z=5$ exceeds the upper critical dimension.

\begin{figure}
\includegraphics[width=3in]{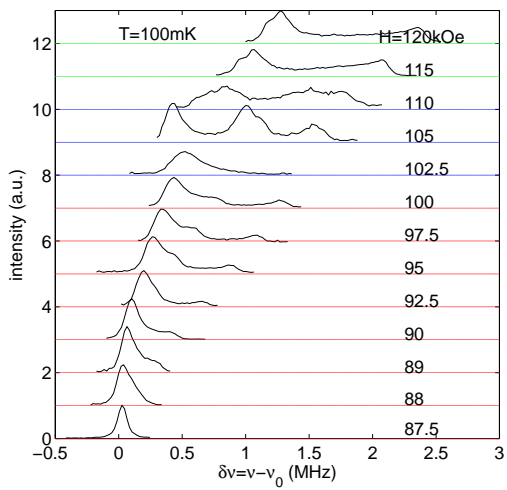}
\caption{(color online) Ba(1) spectra for \HperpC\ close to $H_{c1\perp}$=86.5kOe. Lineshape changes in the range of fields 102kOe$<H<$115kOe are consistent with a line of first order phase transitions dividing phase I and II. }\label{fig:Ba2HperpC}
\end{figure}

Fig.\ref{fig:Ba2HperpC} shows the Ba(2) spectra near $H_{c1\perp}$.
Unlike for \HparC, the spectra are distinctly asymmetric at all fields in the range of
phase II. Still, the spectrum remains relatively simple up to $H\approx$100-105kOe.
The spectrum becomes particularly complicated over a range of fields extending to approximately 115kOe, then simplifies once again, for $H\ge115$kOe. We take this observation as evidence for a line of first order phase transitions dividing phases II and I, consistent with recent neutron scattering experiments carried out independent of this work \cite{Stone:2009}.

It is useful to consider the prediction for the nature of phase II to interpret the significance of the lineshapes. At low temperatures and $H>H_{c1}$, the transverse magnetization oscillates with wavevector $\bf{Q}$ out of the plane, $M_t\parallel\mathbf{c}$. The longitudinal component
is modulated with wavevector $2\bf{Q}$. The behavior is parameterized as
\begin{eqnarray}
S^x_{j \mu}= (-1)^{\mu} \frac{1}{\sqrt{3}} \sin{2 \theta} \cos({\bf r}_j \cdot {\bf Q}), \;\;\;
S^y_{j \mu}=0
\\
S^z_{j \mu}= \frac{1}{4} \pm \frac{1}{4} \sqrt{\cos^2{2 \theta}+\sin^2{2 \theta} \sin^2({\bf r}_j \cdot {\bf Q})}
\label{eq:MzHperpC},
\end{eqnarray}
where j is a dimer index, $\mu=1,2$ labels the two  sites of each dimer, \HperpC defines the z-axis,
and $\theta=[0,\pi]$. The results of model simulations indicate that the lineshapes for $H<$105kOe are qualitatively consistent with phase II, provided that both isotropic and anisotropic hyperfine coupling terms are included \cite{Suh:2009a}. 

The \BaFS\ spectroscopy reported here summarized the behavior around the critical point $H=H_{c1}$ for two directions of ${\bf H}$. For the longitudinal magnetization, the data is well described by including interdimer (near-neighbor) repulsions in the ladder calculation. In particular, the BEC universality class applies for the case \HparC, whereas Ising-like criticality applies to \HperpC. Our data also establishes the presence of incommensurate order parameters. However, some parameters, which are relevant to the analysis of the Ba(2) site hyperfine coupling for \HparC, remain unknown. For example, the shifts of the sites located near Mn triangles of opposite chirality are different in the condensed phase, and it is not clear why this should be the case. A possible explanation could originate with spin-orbit interactions indirectly impacting the hyperfine fields in Ba(2) sites of different chirality.

\acknowledgements The authors acknowledge helpful discussions with T. Giamarchi, O. Sushkov, and
M. Whangbo. This work was supported in part by the NSF under grant numbers DMR-0520552 (SEB) and DMR-0804625 (SEB), DMR-0705087 (IRF), DMR-0602859 (JSB), and by the NNSA of the U.S. DOE at LANL under Contract No. DE-AC52-06NA25396. Research at the NHMFL is supported by the National Science Foundation under grant number DMR-0084173, by the State of Florida, and by the Department of Energy.


\begin{thebibliography}{17}
\expandafter\ifx\csname natexlab\endcsname\relax\def\natexlab#1{#1}\fi
\expandafter\ifx\csname bibnamefont\endcsname\relax
  \def\bibnamefont#1{#1}\fi
\expandafter\ifx\csname bibfnamefont\endcsname\relax
  \def\bibfnamefont#1{#1}\fi
\expandafter\ifx\csname citenamefont\endcsname\relax
  \def\citenamefont#1{#1}\fi
\expandafter\ifx\csname url\endcsname\relax
  \def\url#1{\texttt{#1}}\fi
\expandafter\ifx\csname urlprefix\endcsname\relax\def\urlprefix{URL }\fi
\providecommand{\bibinfo}[2]{#2}
\providecommand{\eprint}[2][]{\url{#2}}

\bibitem[{\citenamefont{Giamarchi et~al.}(2008)\citenamefont{Giamarchi,
  R\"{u}egg, and Tchernyshyov}}]{Giamarchi:2008}
\bibinfo{author}{\bibfnamefont{T.}~\bibnamefont{Giamarchi}},
  \bibinfo{author}{\bibfnamefont{C.}~\bibnamefont{R\"{u}egg}},
  \bibnamefont{and}
  \bibinfo{author}{\bibfnamefont{O.}~\bibnamefont{Tchernyshyov}},
  \bibinfo{journal}{Nature Physics} \textbf{\bibinfo{volume}{4}},
  \bibinfo{pages}{198} (\bibinfo{year}{2008}).

\bibitem[{\citenamefont{Nikuni et~al.}(2000)\citenamefont{Nikuni, Oshikawa,
  Oosawa, and Tanaka}}]{Nikuni:2000}
\bibinfo{author}{\bibfnamefont{T.}~\bibnamefont{Nikuni}},
  \bibinfo{author}{\bibfnamefont{M.}~\bibnamefont{Oshikawa}},
  \bibinfo{author}{\bibfnamefont{A.}~\bibnamefont{Oosawa}}, \bibnamefont{and}
  \bibinfo{author}{\bibfnamefont{H.}~\bibnamefont{Tanaka}},
  \bibinfo{journal}{Phys. Rev. Lett.} \textbf{\bibinfo{volume}{84}},
  \bibinfo{pages}{5868} (\bibinfo{year}{2000}).

\bibitem[{\citenamefont{Jaime et~al.}(2004)\citenamefont{Jaime, Correa,
  Harrison, Batista, Kawashima, Kazuma, Jorge, Stern, Heinmaa, Zvyagin
  et~al.}}]{Jaime:2004}
\bibinfo{author}{\bibfnamefont{M.}~\bibnamefont{Jaime}},
  \bibinfo{author}{\bibfnamefont{V.~F.} \bibnamefont{Correa}},
  \bibinfo{author}{\bibfnamefont{N.}~\bibnamefont{Harrison}},
  \bibinfo{author}{\bibfnamefont{C.~D.} \bibnamefont{Batista}},
  \bibinfo{author}{\bibfnamefont{N.}~\bibnamefont{Kawashima}},
  \bibinfo{author}{\bibfnamefont{Y.}~\bibnamefont{Kazuma}},
  \bibinfo{author}{\bibfnamefont{G.~A.} \bibnamefont{Jorge}},
  \bibinfo{author}{\bibfnamefont{R.}~\bibnamefont{Stern}},
  \bibinfo{author}{\bibfnamefont{I.}~\bibnamefont{Heinmaa}},
  \bibinfo{author}{\bibfnamefont{S.~A.} \bibnamefont{Zvyagin}},
  \bibnamefont{et~al.}, \bibinfo{journal}{Phys. Rev. Lett.}
  \textbf{\bibinfo{volume}{93}}, \bibinfo{eid}{087203} (\bibinfo{year}{2004}).

\bibitem[{\citenamefont{Kodama et~al.}(2002)\citenamefont{Kodama, Takigawa,
  Horvatic, Berthier, Kageyama, Ueda, Miyahara, Becca, and Mila}}]{Kodama:2002}
\bibinfo{author}{\bibfnamefont{K.}~\bibnamefont{Kodama}},
  \bibinfo{author}{\bibfnamefont{M.}~\bibnamefont{Takigawa}},
  \bibinfo{author}{\bibfnamefont{M.}~\bibnamefont{Horvatic}},
  \bibinfo{author}{\bibfnamefont{C.}~\bibnamefont{Berthier}},
  \bibinfo{author}{\bibfnamefont{H.}~\bibnamefont{Kageyama}},
  \bibinfo{author}{\bibfnamefont{Y.}~\bibnamefont{Ueda}},
  \bibinfo{author}{\bibfnamefont{S.}~\bibnamefont{Miyahara}},
  \bibinfo{author}{\bibfnamefont{F.}~\bibnamefont{Becca}}, \bibnamefont{and}
  \bibinfo{author}{\bibfnamefont{F.}~\bibnamefont{Mila}},
  \bibinfo{journal}{Science} \textbf{\bibinfo{volume}{298}},
  \bibinfo{pages}{395} (\bibinfo{year}{2002}).

\bibitem[{\citenamefont{Rice}(2002)}]{Rice:2002}
\bibinfo{author}{\bibfnamefont{T.~M.} \bibnamefont{Rice}},
  \bibinfo{journal}{Science} \textbf{\bibinfo{volume}{298}},
  \bibinfo{pages}{760} (\bibinfo{year}{2002}).

\bibitem[{\citenamefont{Weller and Skinner}(1999)}]{Weller:1999}
\bibinfo{author}{\bibfnamefont{M.~T.} \bibnamefont{Weller}} \bibnamefont{and}
  \bibinfo{author}{\bibfnamefont{S.~J.} \bibnamefont{Skinner}},
  \bibinfo{journal}{Acta Cryst. C} \textbf{\bibinfo{volume}{55}},
  \bibinfo{pages}{154} (\bibinfo{year}{1999}).

\bibitem[{\citenamefont{Samulon et~al.}(2008)\citenamefont{Samulon, Jo,
  Sengupta, Batista, Jaime, Balicas, and Fisher}}]{Samulon:2008}
\bibinfo{author}{\bibfnamefont{E.~C.} \bibnamefont{Samulon}},
  \bibinfo{author}{\bibfnamefont{Y.-J.} \bibnamefont{Jo}},
  \bibinfo{author}{\bibfnamefont{P.}~\bibnamefont{Sengupta}},
  \bibinfo{author}{\bibfnamefont{C.~D.} \bibnamefont{Batista}},
  \bibinfo{author}{\bibfnamefont{M.}~\bibnamefont{Jaime}},
  \bibinfo{author}{\bibfnamefont{L.}~\bibnamefont{Balicas}}, \bibnamefont{and}
  \bibinfo{author}{\bibfnamefont{I.~R.} \bibnamefont{Fisher}},
  \bibinfo{journal}{Phys. Rev. B} \textbf{\bibinfo{volume}{77}},
  \bibinfo{eid}{214441} (\bibinfo{year}{2008}).

\bibitem[{\citenamefont{Uchida et~al.}(2002)\citenamefont{Uchida, Tanaka,
  Mitamura, Ishikawa, and Goto}}]{Uchida:2002}
\bibinfo{author}{\bibfnamefont{M.}~\bibnamefont{Uchida}},
  \bibinfo{author}{\bibfnamefont{H.}~\bibnamefont{Tanaka}},
  \bibinfo{author}{\bibfnamefont{H.}~\bibnamefont{Mitamura}},
  \bibinfo{author}{\bibfnamefont{F.}~\bibnamefont{Ishikawa}}, \bibnamefont{and}
  \bibinfo{author}{\bibfnamefont{T.}~\bibnamefont{Goto}},
  \bibinfo{journal}{Phys. Rev. B} \textbf{\bibinfo{volume}{66}},
  \bibinfo{pages}{054429} (\bibinfo{year}{2002}).

\bibitem[{\citenamefont{Tsujii et~al.}(2005)\citenamefont{Tsujii, Andraka,
  Uchida, Tanaka, and Takano}}]{Tsujii:2005}
\bibinfo{author}{\bibfnamefont{H.}~\bibnamefont{Tsujii}},
  \bibinfo{author}{\bibfnamefont{B.}~\bibnamefont{Andraka}},
  \bibinfo{author}{\bibfnamefont{M.}~\bibnamefont{Uchida}},
  \bibinfo{author}{\bibfnamefont{H.}~\bibnamefont{Tanaka}}, \bibnamefont{and}
  \bibinfo{author}{\bibfnamefont{Y.}~\bibnamefont{Takano}},
  \bibinfo{journal}{Phys. Rev. B} \textbf{\bibinfo{volume}{72}},
  \bibinfo{pages}{214434} (\bibinfo{year}{2005}).

\bibitem[{\citenamefont{Uchida et~al.}(2001)\citenamefont{Uchida, Tanaka,
  Bartashevich, and Goto}}]{Uchida:2001}
\bibinfo{author}{\bibfnamefont{M.}~\bibnamefont{Uchida}},
  \bibinfo{author}{\bibfnamefont{H.}~\bibnamefont{Tanaka}},
  \bibinfo{author}{\bibfnamefont{M.~I.} \bibnamefont{Bartashevich}},
  \bibnamefont{and} \bibinfo{author}{\bibfnamefont{T.}~\bibnamefont{Goto}},
  \bibinfo{journal}{J. Phys. Soc. Japan} \textbf{\bibinfo{volume}{70}},
  \bibinfo{pages}{1790} (\bibinfo{year}{2001}).

\bibitem[{\citenamefont{Singh and Huse}(1992)}]{Singh:1992}
\bibinfo{author}{\bibfnamefont{R.~R.~P.} \bibnamefont{Singh}} \bibnamefont{and}
  \bibinfo{author}{\bibfnamefont{D.~A.} \bibnamefont{Huse}},
  \bibinfo{journal}{Phys. Rev. Lett.} \textbf{\bibinfo{volume}{68}},
  \bibinfo{pages}{1766} (\bibinfo{year}{1992}).

\bibitem[{\citenamefont{Kawamura}(1998)}]{Kawamura:1998}
\bibinfo{author}{\bibfnamefont{H.}~\bibnamefont{Kawamura}},
  \bibinfo{journal}{J. Phys.: Cond. Matter} \textbf{\bibinfo{volume}{10}},
  \bibinfo{pages}{4707} (\bibinfo{year}{1998}).

\bibitem[{\citenamefont{Stone et~al.}(2008)\citenamefont{Stone, Lumsden, Chang,
  Samulon, Batista, and Fisher}}]{Stone:2008a}
\bibinfo{author}{\bibfnamefont{M.~B.} \bibnamefont{Stone}},
  \bibinfo{author}{\bibfnamefont{M.~D.} \bibnamefont{Lumsden}},
  \bibinfo{author}{\bibfnamefont{S.}~\bibnamefont{Chang}},
  \bibinfo{author}{\bibfnamefont{E.~C.} \bibnamefont{Samulon}},
  \bibinfo{author}{\bibfnamefont{C.~D.} \bibnamefont{Batista}},
  \bibnamefont{and} \bibinfo{author}{\bibfnamefont{I.~R.}
  \bibnamefont{Fisher}}, \bibinfo{journal}{Phys. Rev. Lett.}
  \textbf{\bibinfo{volume}{100}}, \bibinfo{eid}{237201} (\bibinfo{year}{2008}).

\bibitem[{\citenamefont{Beliaev}(1958)}]{Beliaev:1958}
\bibinfo{author}{\bibfnamefont{S.~T.} \bibnamefont{Beliaev}},
  \bibinfo{journal}{Sov. Phys. JETP} \textbf{\bibinfo{volume}{7}},
  \bibinfo{pages}{299} (\bibinfo{year}{1958}).

\bibitem[{Hil()}]{Hill:2009}
\bibinfo{note}{S. Hill (private communication)}.

\bibitem[{\citenamefont{Suh et~al.}()\citenamefont{Suh, Brown, Samulon, Fisher,
  and Batista}}]{Suh:2009a}
\bibinfo{author}{\bibfnamefont{S.}~\bibnamefont{Suh}},
  \bibinfo{author}{\bibfnamefont{S.~E.} \bibnamefont{Brown}},
  \bibinfo{author}{\bibfnamefont{E.~C.} \bibnamefont{Samulon}},
  \bibinfo{author}{\bibfnamefont{I.~R.} \bibnamefont{Fisher}},
  \bibnamefont{and} \bibinfo{author}{\bibfnamefont{C.~D.}
  \bibnamefont{Batista}}, \bibinfo{note}{(unpublished)}.

\bibitem[{\citenamefont{Stone}()}]{Stone:2009}
\bibinfo{author}{\bibfnamefont{M.}~\bibnamefont{Stone}},
  \bibinfo{note}{(private communication)}.

\end{thebibliography}

\end{document}